\documentclass[%
 aip,
 amsmath,amssymb,
 reprint,%
]{revtex4-1}

\usepackage{graphicx}
\usepackage{dcolumn}
\usepackage{bm}

\usepackage{color}

\usepackage[utf8]{inputenc}
\usepackage[T1]{fontenc}
\usepackage{mathptmx}

\usepackage{url}
\newcommand{\be}{\begin{eqnarray}}
\newcommand{\ee}{\end{eqnarray}}

\newcommand{\nn}{\nonumber}

\newcommand{\bo}{\boldsymbol}
\newcommand{\lb}{\label}

\begin{document}

\preprint{AIP/123-QED}

\title[Self-learning hybrid Monte Carlo method for isothermal-isobaric ensemble]{Self-learning hybrid Monte Carlo method for isothermal-isobaric ensemble: Application to liquid silica}

\author{Keita Kobayashi}
\email{kobayashi.keita@jaea.go.jp}
\affiliation{CCSE, Japan Atomic Energy Agency, 178-4-4, Wakashiba , Kashiwa, Chiba
 277-0871, Japan}

\author{Yuki Nagai}%
\email{nagai.yuki@jaea.go.jp}
\affiliation{%
CCSE, Japan Atomic Energy Agency, 178-4-4, Wakashiba , Kashiwa, Chiba
 277-0871, Japan}%
 \affiliation{%
Mathematical Science Team, RIKEN Center for Advanced Intelligence Project (AIP), 1-4-1 Nihonbashi, Chuo-ku, Tokyo 103-0027, Japan
 }
 \author{Mitsuhiro Itakura}%
\affiliation{%
CCSE, Japan Atomic Energy Agency, 178-4-4, Wakashiba , Kashiwa, Chiba
 277-0871, Japan}%
\author{Motoyuki Shiga}%
\email{shiga.motoyuki@jaea.go.jp}
\affiliation{%
CCSE, Japan Atomic Energy Agency, 178-4-4, Wakashiba , Kashiwa, Chiba
 277-0871, Japan}%
 
\date{\today}

\begin{abstract}
Self-learning hybrid Monte Carlo (SLHMC) is a first-principles simulation that allows for exact ensemble generation on potential energy surfaces based on density functional theory. 
The statistical sampling can be accelerated with the assistance of smart trial moves by machine learning potentials. 
In the first report (Nagai, {\it et al}. Phys. Rev. B 102, 041124(R) (2020)), the SLHMC approach was introduced for the simplest case of canonical sampling. We herein extend this idea to isothermal-isobaric ensembles to enable general applications for soft materials and liquids with large volume fluctuation.
As a demonstration, the isothermal-isobaric SLHMC method was used to study the vibrational structure of liquid silica at temperatures close to the melting point, whereby the slow diffusive motion is beyond the time scale of first-principles molecular dynamics. 
It was found that the static structure factor thus computed from first-principles agrees quite well with the high-energy X-ray data.
\end{abstract}

\maketitle

\section{Introduction}
Atomistic simulations have become an essential tool for studying physical and chemical properties of materials. 
Molecular dynamics (MD) or Markov chain Monte Carlo (MCMC) simulations rely on the quality of potential energy surface (PES) describing the system of interest, and the first-principles density functional theory (DFT) \cite{PhysRev.140.A1133} is a standard choice.
However, statistical sampling with the DFT-MD method often requires large computational effort, especially for soft materials.
As an alternative to DFT-MD, it would be useful to develop an MCMC method that accelerates the generation of statistical ensemble on the DFT-PES.
For this purpose we recently proposed the self-learning hybrid Monte Carlo (SLHMC) method \cite{Nagai2020}.

SLHMC is a hybrid Monte Carlo (HMC) technique \cite{PhysRevD.35.2531,DUANE1987216,PhysRevB.45.679,Shinoda2004,Nakayama2009} combined with an auxiliary use of machine learning potential (MLP) \cite{BPNN1,GAP1, BPNN2,GAP2,Behler2016}.
In this method, the atomic force from MLP is used to design efficient HMC trial moves.
Yet the acceptance is determined such that the HMC sampling is done exactly on a DFT-PES.
The MLP can be trained to imitate the DFT potential during the sampling process, which helps increase the HMC acceptance ratio at larger step intervals.
In our first report \cite{Nagai2020}, the SLHMC method was introduced for the case of canonical ensemble.
The applications were then demonstrated for solid systems, such as the radial distribution functions of $\alpha$-quartz SiO${}_{2}$ and the phonon density of states of superconductor YNi${}_{2}$B${}_{2}$C.
However, the simulations in the canonical ensemble are often not suitable for liquids and soft materials where the volume fluctuation is involved.
In this paper, we extend the idea of SLHMC for the simulations in the isothermal-isobaric ensemble to take account of the effects of volume fluctuation.
The method developed is then used to study the structural properties of liquid silica.
Liquid silica is not only of fundamental importance in geoscience, but also of industrial interest as a glass-forming material.
Despite its significance, structural studies in experiment were limited due to the high melting temperature \cite{Mei2007}.
In those cases, atomistic simulations are able to provide predictive data.
MD simulations of liquid silica with various empirical force fields \cite{Vashishta1990,Kob_1999,Ryltsev2013,Geske2016} and DFT-MD \cite{Sarnthein1995_prl,Sarnthein1995_prb,PhysRevB.86.054104} have been reported so far.
Previous DFT-MD simulations were conducted at a temperature well above the melting point since the diffusion of liquid silica is extremely slow due to the strong covalent Si-O bonds. 
The reason is presumably that it has been difficult to capture the slow diffusion at a realistic temperature in short DFT-MD runs. 
The MLP of silica \cite{Li2018,Balyakin2020} may enable longer MD runs.
However, the development of robust MLP is not easy since it requires a fine-tuning of hyperparameters and a proper selection of reference data \cite{aenet,Miksch2021}.
Validation tests are also required for MLP to reproduce various physical quantities such as lattice parameters, elastic properties, radial distribution functions, and diffusion constants.
An advantage of SLHMC is that, in principle, the generation of exact statistical ensemble on DFT-PES is always guaranteed, even if the MLP employed does not perfectly imitate the DFT potential.
Importantly, the efficient sampling of SLHMC should extend the possibility to explore soft materials that are beyond the range of standard DFT-MD, while the results are in the same quality as (a longer run of) DFT-MD.

This paper is organized as follows.
First, we introduce the theory and method of SLHMC with isothermal-isobaric ensemble.
Next, we demonstrate the efficiency and accuracy of the SLHMC compared with the DFT-MD simulation for liquid silica.
Finally, we compare the computed results to the experimental ones.

\section{theory}
We consider the NPT ensemble for a system of $N$ atoms
 contained in a flexible parallel-piped hexagonal box that allows
 for anisotropic fluctuation \cite{Tuckerman}.
The partition function is given as
\begin{eqnarray}
 \Xi = \int_0^\infty d\bo{h} \frac{Z(N,V,T)}{V^2} \exp(-\beta P_{\rm ext}^{} V),
\end{eqnarray}
where $\bo{h} =(\bo{a},\bo{b},\bo{c})$ is the ($3\times 3$) box matrix,
 $V={\rm det}[\bo{h}]$ is the box volume, $P_{\rm ext}^{}$ is the
 external pressure, and
\begin{eqnarray}
 Z(N,V,T) = \left(\frac{1}{2\pi\hbar}\right)^{3N}
 \int e^{-\beta H_{\rm X}^{}(\{\bo{r}\},\{\bo{p}\})} d\bo{r} d\bo{p}\,. \lb{eq:partition}
\end{eqnarray}
 is the partition function of the canonical ensemble.
In Eq.\eqref{eq:partition}, $\beta =1/k_{\rm B}T$ and $\hbar=h/2\pi$, where $k_{\rm B}$ and $h$ are Boltzmann and Planck constants, respectively.
For convenience, the index ${\rm X}$ expresses either DFT or MLP.
The system Hamiltonian is a function of the set of atomic
 positions $\{\bo{r}\}=(\bo{r}_1^{},\cdots\bo{r}_N^{})$ and the set of
 atomic momenta and $\{\bo{p}\}=(\bo{p}_1^{},\cdots\bo{p}_N^{})$ as
\begin{eqnarray}
H_{\rm X}^{}(\{\bo{r}\},\{\bo{p}\})
 = \sum_{i=1}^N \frac{(\bo{p}_i^{})^{2}}{2m_i^{}} + \phi_{\rm X}^{}(\{\bo{r}\},\bo{h})\,,
\end{eqnarray}
 where $m_i^{}$ is the atomic mass and $\phi_{\rm X}^{}(\bo{r},\bo{h})$
 is the potential function.


Now we introduce a ($3\times 3$) momentum matrix $\bo{p}_{\rm g}$
 conjugate to $\bo{h}$.
Using notations $\bo{R} = (\{\bo{r}\},\bo{h})$ and
 $\bo{P} = (\{\bo{p}\},\bo{p}_{\rm g}^{})$ to express
 the positions and the momenta in the extended phase space,
 we introduce a pseudo-Hamiltonian of the form
\begin{eqnarray}
\mathcal{H}_{\rm X}^{}(\bo{R},\bo{P})
 = H_{\rm X}^{}(\{\bo{r}\},\{\bo{p}\}) + 
 \frac{\mathrm{Tr}[\bo{p}_{\rm g}^{t}\bo{p}_{\rm g}^{}]}{2W_{\rm g}^{}}
 +P_{\rm ext}^{}{\rm det}[\bo{h}]\,,  \lb{hamiltonian}
\end{eqnarray}
where the second term in the rhs represents the kinetic energy of the barostat whose mass is $W_{\rm g}^{}$.
It is known that $\mathcal{H}_{\rm X}^{}(\bo{R},\bo{P})$
 is conserved for the set of following equations of motion;
for the case of the ${\rm X = MLP}$,
\begin{eqnarray}
\lb{eom_r} 
\dot{\bo{r}}_{i}^{} &=& \frac{\bo{p}_{i}^{}}{m_{i}^{}} + \frac{\bo{p}_{\rm g}^{}}{W_{\rm g}^{}}\bo{r}_{i}^{} \,, \\
 \lb{eom_cell}
\dot{\bo{h}} &=& \frac{\bo{p}_{\rm g}^{}}{W_{\rm g}^{}}\bo{h}  \,,\\
\lb{eom_p} 
\dot{\bo{p}}_{i}^{} &=& -\frac{\partial\phi_{\rm MLP}^{} }{\partial \bo{r}_{i}^{}}
-\frac{\bo{p}_{\rm g}^{}}{W_{\rm g}^{}}\bo{p}_{i}^{}
-\frac{1}{3N}\frac{ \mathrm{Tr}[\bo{p}_{\rm g}^{}] }{ W_{\rm g}^{} }\bo{p}_{i}^{} \,,\\
\lb{eom_cellp} 
\dot{\bo{p}}_{\rm g}^{} &=& \det[\bo{h}](\bo{{\rm P}}_{\rm int}^{}-\bo{{\rm I}}P_{\rm ext}^{})
+\frac{1}{3N}\sum_{i=1}^{N}\frac{\bo{p}_{i}^{}}{m_{i^{}}}\bo{{\rm I}^{}} \,,
\end{eqnarray}
where $\bo{{\rm I}}$ is the unit matrix.
The internal pressure tensor $\bo{{\rm P}}_{{\rm int}}^{}$ is given by
\be
(\bo{{\rm P}}_{{\rm int}}^{})_{\alpha\beta}^{}&=&\frac{1}{\det[\bo{h}]}
\left[
\sum_{i=1}^{N}\frac{(\bo{p}_i^{})_\alpha^{}(\bo{p}_i^{})_\beta^{}}{m_{i}^{}}
-\frac{\partial\phi_{\rm MLP}^{} }{\partial (\bo{r}_i^{})_\alpha^{}} (\bo{r}_i^{})_\beta^{}\right] \nn\\
&&
-\frac{1}{\det[\bo{h}]}\sum_{\gamma=x,y,z}\frac{\partial \phi_{\rm MLP}^{}}{\partial (\bo{h})_{\alpha\gamma}^{}}
(\bo{h})_{\gamma\beta}^{}\,.
\ee
In the phase space,
 $\bo{\Gamma} = \left( \{{\bf R}\},\{{\bf P}\} \right)$,
 the energy conservation can be checked from Equations 
\eqref{eom_r}--\eqref{eom_cellp} as
\begin{eqnarray}
 \dot{\mathcal{H}}_{\rm MLP}^{}(\bo{R},\bo{P})
 = \frac{\partial\mathcal{H}}{\partial \bo{\Gamma}}\dot{\bo{\Gamma}}= 0\,. \lb{conserve}
\end{eqnarray}
Meanwhile the change of Jacobian of the phase space $\mathcal{J}(\bo{\Gamma})$
 along the trajectory is subject to the relation
\begin{eqnarray}
 \frac{\dot{\mathcal{J}}(\bo{\Gamma})}{\mathcal{J}(\bo{\Gamma})}
 = - \frac{\partial\dot{\bo{\Gamma}}}
 {\partial\bo{\Gamma}}\,. \lb{J1}
\end{eqnarray}
From Equations \eqref{eom_r}--\eqref{eom_cellp} we obtain
\begin{eqnarray}
 \frac{\partial\dot{\boldsymbol{\Gamma}}}
 {\partial\boldsymbol{\Gamma}} =
 2 {\rm Tr} \left( \frac{\bo{p}_g^{}}{W_g^{}} \right)
 =
 2{\rm Tr} \left( \dot{\bo{h}}\bo{h}^{-1} \right)
 =
 2\frac{d}{dt} \ln \{{\rm det}(\bo{h})\}\,, \lb{J2}
\end{eqnarray}
and so the solution to Equation \eqref{J1} is
\begin{eqnarray}
 \mathcal{J}(\boldsymbol{\Gamma}) \propto
 \left\{{\rm det}(\bo{h})\right\}^{-2}\,. \lb{J3}
\end{eqnarray}
%


%
Now, the target distribution function is the NPT ensemble based on the DFT potential,
\begin{eqnarray}
 f(\bo{R},\bo{P}) \propto \{{\rm det}[\bo{h}]\}^{-2}
 \exp(-\beta \mathcal{H}_{\rm DFT}^{}(\bo{R},\bo{P}))\,.\lb{prob}
\end{eqnarray}
To generate this distribution, the HMC algorithm should
be designed to obey the detailed balance condition 
with respect to the points in the configuration space, $\bo{R}$ and $\bo{R}^\prime$,
\begin{eqnarray}
 &&
 \int d\bo{P}^{} d\bo{P}^\prime
 T(\bo{R}^\prime,\bo{P}^\prime|\bo{R},\bo{P})
 A(\bo{R}^\prime,\bo{P}^\prime|\bo{R},\bo{P})
 f(\bo{R},\bo{P}) =
 \nonumber \\ &&
 \int d\bo{P} d\bo{P}^{\prime}
 T(\bo{R},\bo{P}|\bo{R}^\prime,\bo{P}^\prime)
 A(\bo{R},\bo{P}|\bo{R}^\prime,\bo{P}^\prime)
 f(\bo{R}^\prime,\bo{P}^\prime),\lb{balance}
\end{eqnarray}
 where
 $T(\bo{R}^\prime,\bo{P}^\prime|\bo{R},\bo{P})$ and
 $A(\bo{R}^\prime,\bo{P}^\prime|\bo{R},\bo{P})$ are
 the trial and acceptance probabilities, respectively, in the MCMC algorithm
 with respect to the move from $\bo{R},\bo{P}$ to $\bo{R}^\prime,\bo{P}^\prime$.
Since the trial moves from the equations of motions \eqref{eom_r}--\eqref{eom_cellp} are deterministic and reversible,
 the ratio of trial probabilities $T$ of the forward and backward moves is proportional
 to the ratio of Jacobian $\mathcal{J}$ of the end points of the move, and thus,
\begin{eqnarray}
 \frac{T(\bo{R}^\prime,\bo{P}^\prime|\bo{R},\bo{P})}
 {T(\bo{R},\bo{P}|\bo{R}^\prime,\bo{P}^\prime)}
 = \frac{\{{\rm det}[\bo{h}]\}^2}{\{{\rm det}[\bo{h}^\prime]\}^2}\,.\lb{PR}
\end{eqnarray}
Equation \eqref{balance} is satisfied then by setting the acceptance probability as
\begin{eqnarray}
 &&
 A(\bo{R}^\prime,\bo{P}^\prime|\bo{R},\bo{P})
 =
 \nonumber \\ &&
 \min\left\{1,e^{-\beta( \mathcal{H}_{\rm DFT}^{}(\bo{R}^\prime,\bo{P}^\prime)
 - \mathcal{H}_{\rm DFT}^{}(\bo{R},\bo{P}))}\right\}\,. \lb{AC1}
\end{eqnarray}
This could be proved by the substitution of Equations \eqref{prob}, \eqref{PR} and \eqref{AC1}
 into Equation \eqref{balance} and noting the time reversibility,
 $T(\bo{R}^\prime,\bo{P}^\prime|\bo{R},\bo{P})
 = T(\bo{R},-\bo{P}|\bo{R}^\prime,-\bo{P}^\prime)$.
With a small enough step size, the total energy should conserve,
 $\mathcal{H}_{\rm MLP}^{}(\bo{R}^\prime,\bo{P}^\prime)
 = \mathcal{H}_{\rm MLP}^{}(\bo{R},\bo{P})$, so that
 Equation \eqref{AC1} finally reduces into a simple form,
\begin{eqnarray}
 &&A(\bo{R}^\prime,\bo{P}^\prime|\bo{R},\bo{P})
 = \nn\\
 &&\min\left\{1,e^{-\beta( \Delta\phi(\{\bo{r}^\prime\},\bo{h}^\prime)
 - \Delta\phi(\{\bo{r}\},\bo{h}))}\right\}\,, \lb{AC2}
\end{eqnarray}
 where
\begin{eqnarray}
 \Delta\phi(\{\bo{r}\},\bo{h}) &\equiv&
 \phi_{\rm DFT}^{}(\{\bo{r}\},\bo{h}) - \phi_{\rm MLP}^{}(\{\bo{r}\},\bo{h})
\end{eqnarray}
 is the difference between the DFT and MLP potentials.
Thus, the change in the energy $\Delta\phi$ on trial moves,
 relative to the thermal energy $\beta^{-1}$,
 is the factor that directly affects the acceptance.
We note that the acceptance becomes 100\% in the ideal case where the DFT and MLP potentials are perfectly identical.


The SLHMC algorithm is summarized as follows.
At each MCMC step, all the atomic and barostat momenta, $\bo{P}$, are
 randomized according to the Maxwell-Boltzmann distribution.
Then a trial move of $(\bo{R},\bo{P})$ is generated as
 a short-length MD trajectory with respect to
 the solution of Eqs.\eqref{eom_r}--\eqref{eom_cellp}  using the MLP.
Finally, the acceptance is judged by Eq.\eqref{AC1} 
where the DFT potential calculation is required.
The MCMC interval (i.e., the MD step length)
 of the trial move is a measure of the computational efficiency of SLHMC.
This is an adjustable parameter that can be extended until it reaches two possible origins of
 slowdown.
One is the saturation of the acceptance ratio, which is heavily dependent on
 the quality of the MLP, as can been seen from Eq.\eqref{AC2}.
For this reason the MLP could be retrained using the data during
 the sampling process.
The other is a rare case where the computational effort
 of MLP forces and stress tensors become greater than DFT calculations.

\section{Method}
 In this paper, the SLHMC simulations of liquid silica were conducted in the NPT ensemble. 
The SLHMC was implemented in \texttt{PIMD} software\cite{pimd,RUIZBARRAGAN2016130}, 
which supports the interface to both Vienna Ab initio Simulation Package (\texttt{VASP})\,\cite{VASP1,VASP2} 
and Atomic Energy Network (\texttt{aenet}) software\,\cite{aenet,Cooper2020,PhysRevB.96.014112,aenet_code}. 

\subsection{DFT calculation}
The \texttt{VASP} software was used for the DFT calculation based on the projector-augmented wave (PAW) method\,\cite{PAW}. 
The cutoff energy was 500 eV and only the $\Gamma$ point was chosen. 
For the exchange-correlation functional, the generalized gradient approximation of Perdew-Burke-Ernzerhof (GGA-PBE)\,\cite{PBE} was used.

As a reference, DFT-MD simulations were carried out for liquid silica with 72 atoms.
The results were used for the comparison with the SLHMC simulations.
The combination of Langevin thermostat and Parrinello-Rahman barostat 
 were adopted to generate the NPT ensemble. 
The friction coefficients for Langevin dynamics were set as 10 and 20 ps${}^{-1}$ 
 for atomic and cell motions, respectively, and
 the mass of thermobarostat was 1000 atomic mass unit.
The DFT-MD simulations were conducted for 80 ps each at temperatures
 2500, 3000, 3500, 4000, and 4500 K
 with the step size of 1 fs.

\subsection{Machine learning potential}
For the MLP, we adopted the artificial neural network (ANN) potentials of the Behler-Parrinello type \cite{BPNN1, BPNN2}, which was created by the \texttt{aenet} software.
In the ANN method, a local environment of each atom within a cutoff radius $R_{\rm c}$ is encoded in the descriptor vectors, $\bo{G}$. 
The symmetry functions as the descriptors of the radial distances and the angles between atoms are defined by
\be
&&G_i^{\rm (R)} = \sum_j \mathrm{e}^{- \eta^{\rm (R)} \left( R_{ij} - R_\mathrm{s} \right)^2} f_\mathrm{c} \! \left( R_{ij} \right) ,\\
&&G_i^{\rm (A)} = 2^{1 - \zeta} \sum_{j,k \neq i} \sum_{j < k} \left( 1 + \lambda \cos{\theta_{ijk}} \right)^{\! \zeta}  \nonumber \\
&&  \times
 \mathrm{e}^{- \eta^{\rm (A)} \left( R_{ij}^2 + R_{ik}^2 + R_{jk}^2 \right)}
 f_\mathrm{c} \! \left( R_{ij} \right) f_\mathrm{c} \! \left( R_{ik} \right) f_\mathrm{c} \! \left( R_{jk} \right) ,
\ee
with the cutoff function
\be
f_\mathrm{c} (R) =
  \begin{cases}
    \displaystyle
    \frac{1}{2} \left[ \cos \! \left( \frac{\pi R}{R_\mathrm{c}} \right) +1\right]& \left( R \le R_\mathrm{c} \right) \\
    0 & \left( R > R_\mathrm{c} \right)
  \end{cases} \, ,
\ee
where $R_{ij}$ is the distance between atoms $i$ and $j$, 
and $\theta_{ijk}$ is the angle among atoms $i$, $j$ and $k$.
The parameters in the symmetry functions were taken as 
$\eta^{\rm (R)}$=(0.0032 \AA$^{-2}$, \, 0.0357 \AA$^{-2}$, \, 0.0714 \AA$^{-2}$, \, 0.1250 \AA$^{-2}$, 0.2142 \AA$^{-2}$, \, 0.3571 \AA$^{-2}$, \, 0.7142 \AA$^{-2}$, \, 1.4284 \AA$^{-2}$), 
$R_\mathrm{s}=0$ \AA, 
$\eta^{\rm (A)}$=(0.0004  \AA$^{-2}$,  0.02857 \AA$^{-2}$, 0.0893 \AA$^{-2}$), 
$\lambda=(1, \, -1)$,  
$\zeta=(1, \, 2, \, 4)$, and 
$R_{\rm c} = 6.5$ \AA. 
For the ANN architecture, we used 2 hidden layers, 10 nodes in each hidden layer, and the hyperbolic tangent activation function. 
With the atomic energies obtained from the ANN output, $e(\bo{G}_{i})$,
the MLP is represented as
\be
\phi_{\rm MLP}(\{\bo{r}\},\bo{h}) = \sum_{i=1}^{N}e(\bo{G}_{i})\,. \lb{MLP}
\ee
The initial MLP was created by a training set of 8600 structures taken randomly from the DFT-MD simulations.
The L-BFGS method was used for the ANN optimization.

The sufficient accuracy of MLP to conduct the SLHMC is that the difference between DFT and MLP energies settles within a few times the magnitude of the thermal fluctuation, $k_{\rm B}T$.
An important factor for conducting efficient SLHMC simulations is that the ANN architecture of MLP has enough flexibility to imitate the DFT-PES to achieve this. 
Being that satisfied, the SLHMC is designed to improve the accuracy of MLP by sequentially adding the sampled DFT data into training dataset, in principle.

\subsection{SLHMC simulation}
In SLHMC trial moves the equations of motions of
Eqs.\eqref{eom_r}-\eqref{eom_cellp} were numerically integrated
by the reversible reference system propagator algorithm (RESPA).
The force and stress tensors with respect to the MLP were computed by the derivatives of Eq.\eqref{MLP}. 
The step size was chosen to be 0.25 fs with which the Hamiltonian of Eq.\eqref{hamiltonian} conserves well between the MCMC intervals. 
%
%
%
In our calculation, the MCMC interval, $dt_{\rm MC}$, was automatically adjusted between 16 and 256 fs. 
When the acceptance ratio of the last 50 MCMC steps, $P_{\rm ac}^{\rm (50)}$, is more than 
20\%  (less than 5\%), the MCMC interval $dt_{\rm MC}$ is increased (decreased) as 
$dt_{\rm MC} \to 2dt_{\rm MC}$ ($dt_{\rm MC} \to dt_{\rm MC}/2$) every 50 MCMC steps within the range, $16\le dt_{\rm MC}\le 256$. 
On the other hand, if the acceptance ratio of the last 50 MCMC steps is within the range $5 \% \le P_{\rm ac}^{\rm (50)} \le 20 \%$, the MCMC interval $dt_{\rm MC}$ is unchanged.
We conducted the SLHMC simulations with the number of MCMC steps (20000) for liquid silica with 72 and 216 atoms, respectively, at ambient pressure. 
The MLP was retrained every 1000 MCMC steps by adding the sampled DFT data into the training set.

\section{Results}

\begin{table}[htb]
\caption{Acceptance ratio $P_{\rm ac}$ and mean MCMC interval $dt_{\rm m}$ of SLHMC 
for liquid silica with 72 and 216 atoms. 
N mean the number of atoms. 
$t_{\rm eff}$ is the product of $P_{\rm ac}$, $dt_{\rm m}$, and the number of MCMC steps (20000). }
\label{efficiency}
  \begin{tabular}{llllll} \hline
   N &temperature (K)& $P_{\rm ac}$ (\%)  & $dt_{\rm m}$ (fs)  & $t_{\rm eff}$ (ps) \\ \hline
   &2500 & 34.1  & 242  &  1646 \\
  72 &3000 & 31.0  & 217 &  1343 \\
   &3500 & 33.8  & 235 &  1590 \\ 
   & &  &   &   \\
   &2373 & 27.2  & 210  &  1142 \\
  216 &3000 & 23.8  & 188 &  893 \\
   &3500 & 17.4 &  111 &  388 \\ \hline
  \end{tabular}
\end{table}

First, we show the computational efficiency of SLHMC.
The acceptance ratios $P_{\rm ac}$ of SLHMC were from 17 to 34\% as summarized in Table \ref{efficiency}.
The MCMC interval was automatically adjusted during the SLHMC simulation as mentioned in the method section. 
The resulting mean MCMC intervals $dt_{\rm m}$ were listed in Table \ref{efficiency}. 
It is shown that the SLHMC simulations were successfully performed with long intervals $dt_{\rm m}$.
The performance, which is characterized by the values of $P_{\rm ac}$ and $dt_{\rm m}$, can be deteriorated as the system is larger, reflecting the acceptance probability of Eq.\eqref{AC1}. 
In fact, it was found that $P_{\rm ac}$ and $dt_{\rm m}$ were lower for the simulations of 216 atoms than those of 72 atoms. 
It was also found that $P_{\rm ac}$ and $dt_{\rm m}$ were lower as the temperature is higher.
However, the performance cannot be ascribed to the SLHMC method itself. 
According to Eq.\eqref{AC2}, it is because of the quality of MLP, which is deteriorated for temperatures higher than 3500 K (see Appendix A).
We define the effective simulation time, $t_{\rm eff}$, as the product of $P_{\rm ac}$, $dt_{\rm m}$, and the number of MCMC steps (20000). 
The configurations obtained from an SLHMC simulation within $t_{\rm eff}$ are comparable with those obtained from the same length of a DFT-MD simulation.
The $t_{\rm eff}$ values summarized in Table I show that the SLHMC simulations could reach beyond the sub-ns scale which is the conventional range of DFT-MD.

\begin{figure}[h]
\begin{center}
\includegraphics[width=1.0\linewidth]{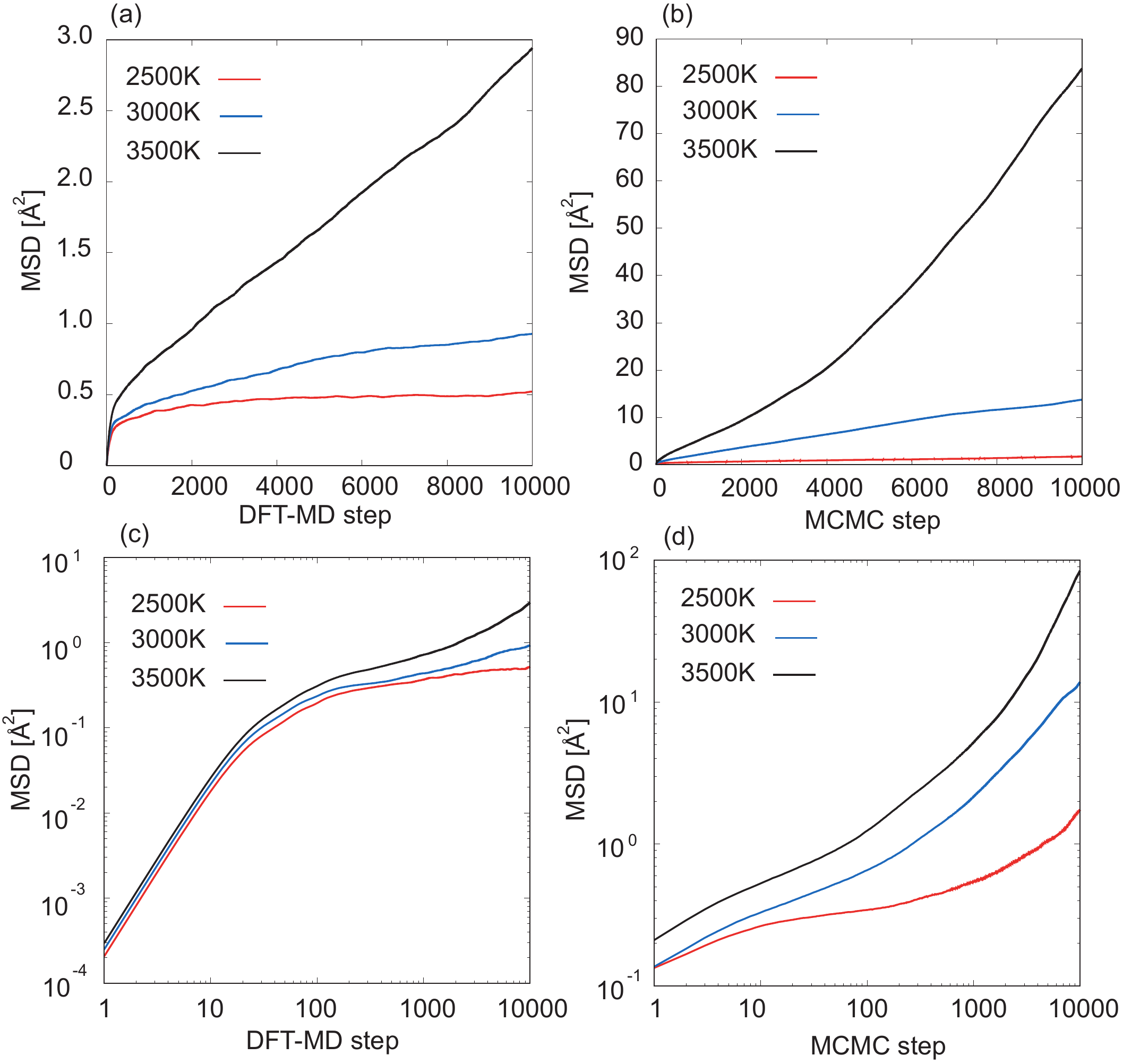}
\end{center}
\caption{
(a) and (b):
Mean square displacement of Si atom for liquid silica obtained by DFT-MD and SLHMC at 2500, 3000, and 3500 K.
The horizontal axis is the number of DFT-MD and MCMC steps, which are equivalent to the number of DFT calculations.  
(c) and (d) are the log-log plot of (a) and (b), respectively.
}
\label{Fig1}
\end{figure}

\begin{figure}[h]
\begin{center}
\includegraphics[width=0.9\linewidth]{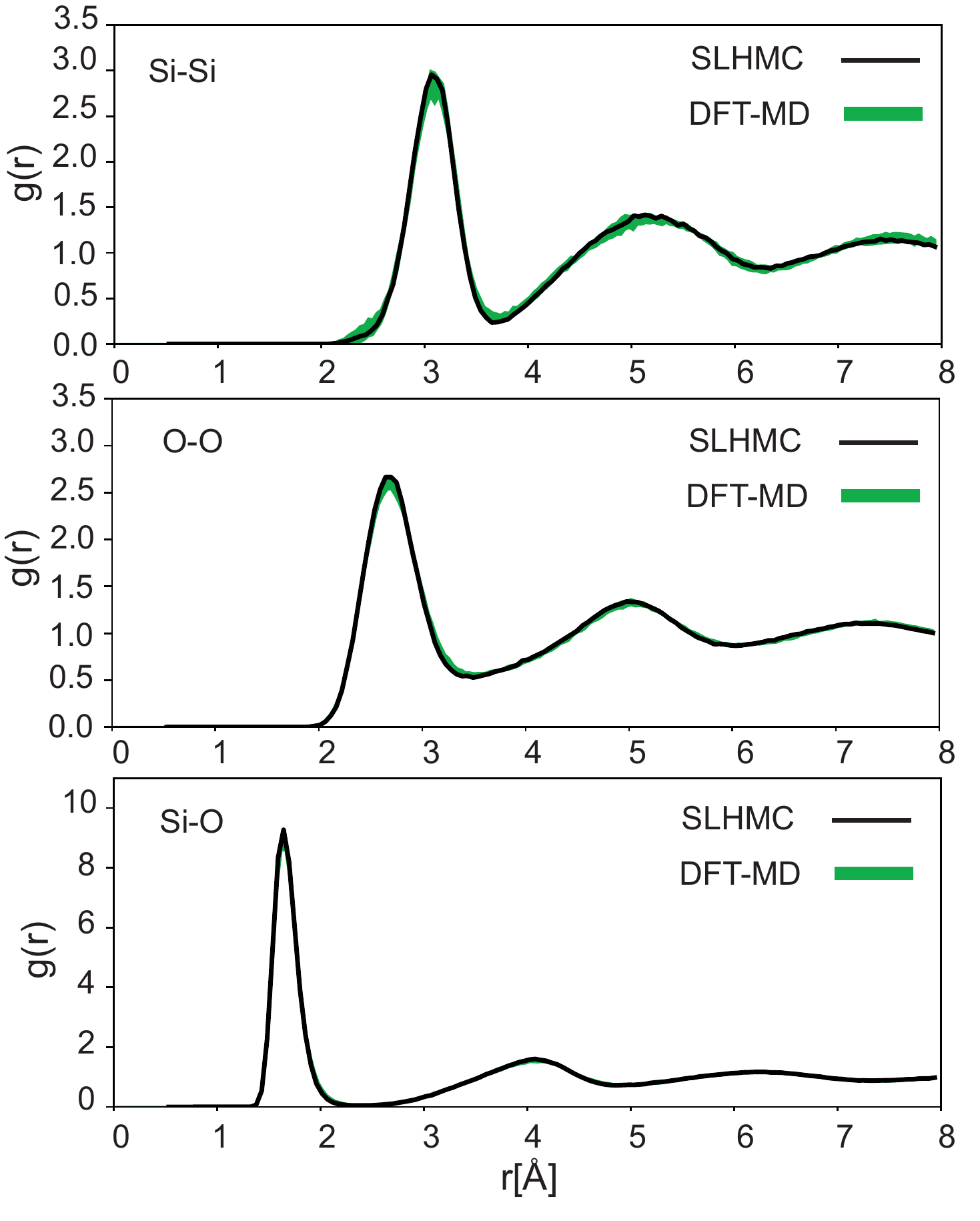}
\end{center}
\caption{
Radial distribution functions $g(r)$ of liquid silica with 72 atoms at 3500 K. 
Black lines are the results obtained by SLHMC. 
In DFT-MD, we divide 80 ps MD trajectory into four blocks and $g(r)$ are calculated in each block. 
$g(r)$ obtained by DFT-MD are shown along with the error bars in green. 
}
\label{Fig2}
\end{figure}

Next, we show the SLHMC simulation results for liquid silica with 72 atoms and compare that with the results obtained by DFT-MD simulation.
Figure \ref{Fig1} shows the mean square displacement (MSD) of Si atoms 
obtained by the SLHMC and DFT-MD.
In the SLHMC simulation, the computational bottleneck is the DFT calculation since  
the contribution from the generation of a trial move is relatively small. 
In SLHMC, the time spent on the MLP training is heavily dependent on the training frequency and the epochs of optimization, but it is generally much smaller than the DFT calculation (See Figure S1 in the Supplemental Material of Reference \cite{Nagai2020}).
To compare the numerical efficiencies of the SLHMC and DFT-MD, the MSD results are shown with respect to the number of DFT calculations in the respective simulations.
As shown in Fig.\ref{Fig1}(a) and (c), the development of MSD obtained by DFT-MD was slow 
due to the presence of the strong covalent Si-O bonds. 
The MSD of typical glass-forming liquid such as silica has three dynamical regimes \cite{Kob_1999,Ryltsev2013,Geske2016}: 
the ballistic regime at short times, 
the plateau at intermediate times, 
and the diffusive regime at long times. 
In the DFT-MD simulation, the diffusive regime in the MSDs was not clearly detectable until 3500 K (see Fig.\ref{Fig1}(c)), 
which is well above the experimental melting point, 1983 K\, \cite{noauthororeditor2007handbook}.  
On the other hand, as expected from the computational efficiency of the SLHMC, the MSDs computed by SLHMC developed much faster than the DFT-MD results with the same number of DFT calculations as shown in Fig.\ref{Fig1}(b) and (d). 
The diffusive regimes of the MSDs were clearly observed at all temperatures in the SLHMC simulation. 
Figure \ref{Fig2} also show the radial distribution functions $g(r)$ at 3500 K. 
The radial distribution functions obtained by SLHMC were in good agreement with the results of 80 ps time DFT-MD run, although the latter had a larger statistical error.
The SLHMC simulation successfully performed the efficient statistical sampling on DFT-PES with a small number of DFT calculations.

\begin{figure}[h]
\begin{center}
\includegraphics[width=1.0\linewidth]{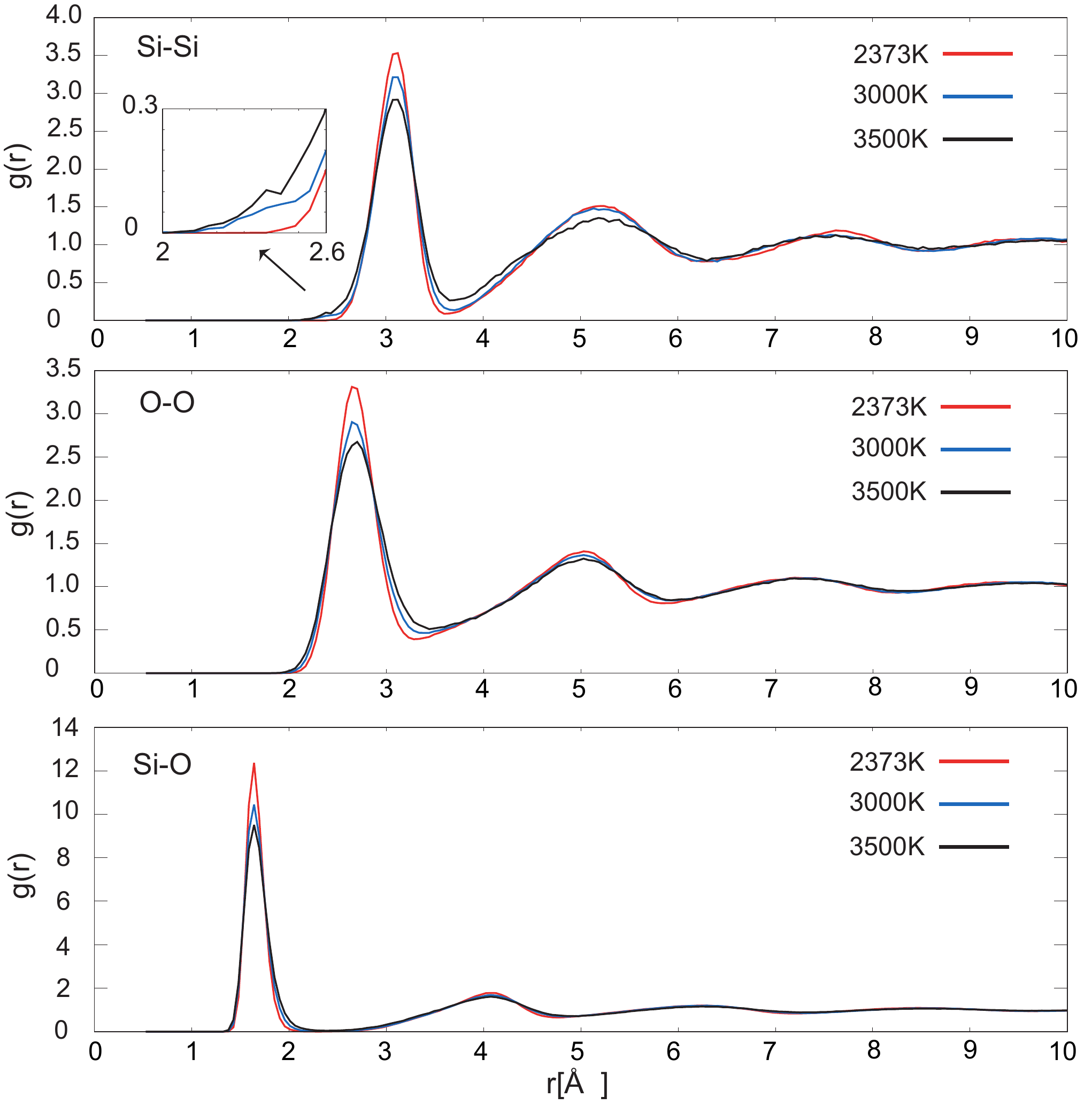}
\end{center}
\caption{
Radial distribution functions $g(r)$ obtained by SLHMC for liquid silica with 216 atoms. 
}
\label{Fig3}
\end{figure}

\begin{figure}[h]
\begin{center}
\includegraphics[width=1.0\linewidth]{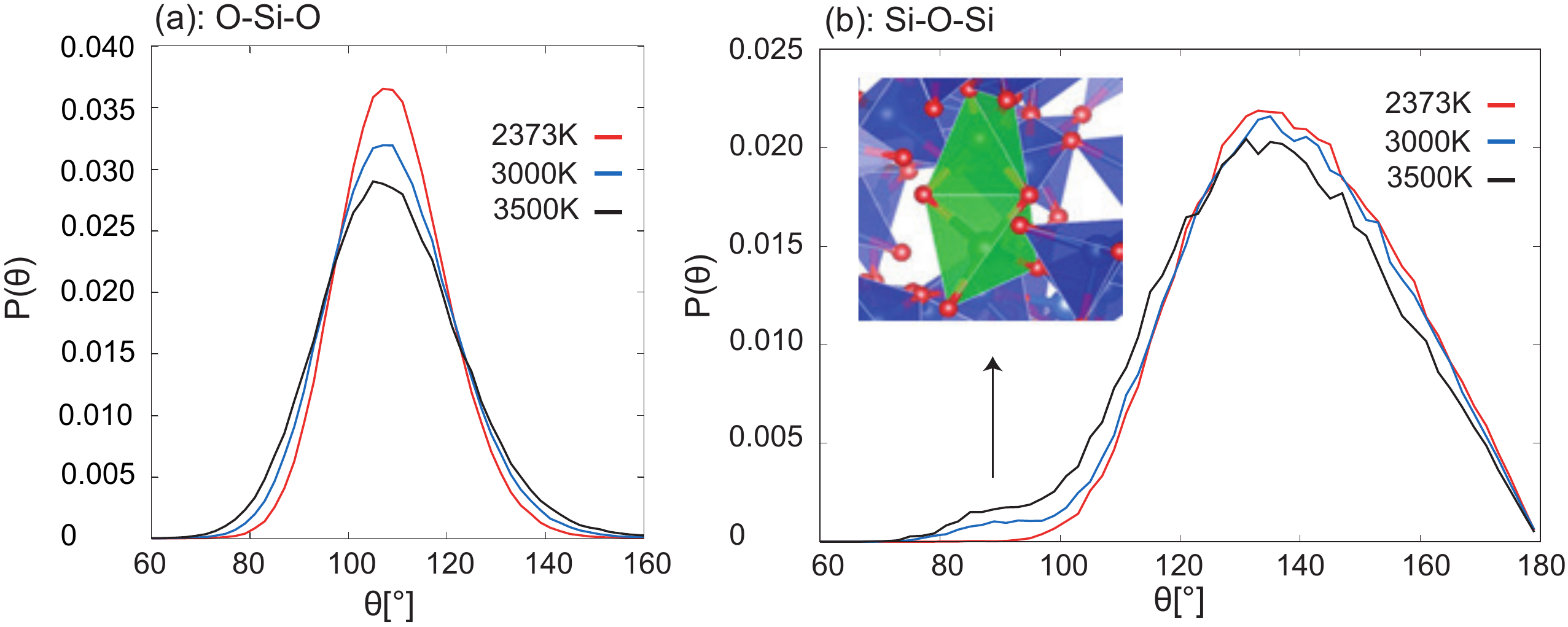}
\end{center}
\caption{
(a) and (b): Bond angle distributions $P(\theta)$ obtained by SLHMC for O-Si-O and Si-O-Si angles. 
The inset of Fig.(b) shows the edge sharing tetrahedrons (green objects) that contribute to the Si-O-Si angle distribution around 90${}^{\circ}$.
}
\label{Fig4}
\end{figure}

\begin{table}[htb]
\caption{The density, bond distance and average bond angle obtained by SLHMC at 2373, 3000, and 3500 K. The bond distances are defined as the first peak position of radial distribution function.}
\label{tab1}
  \begin{tabular}{lllll} \hline
    temperature (K) &      & 2373 & 3000 & 3500 \\ \hline 
  density (\AA${}^{-3}$)    &  & 0.063  & 0.063  &  0.062 \\
  & & &  & \\
  bond distance (\AA)    &Si-Si & 3.13  &  3.13 &  3.13 \\
   &O-O & 2.64 & 2.64 &  2.69\\
    &Si-O  & 1.63 & 1.63 &  1.63\\ 
    & & &  & \\
 average bond angle (${}^{\circ}$)  &O-Si-O  &   109.11 & 108.99  & 108.96\\
    &Si-O-Si & 139.3 & 138.0 & 135.4 \\ \hline
  \end{tabular}
\end{table}

Finally, we show the results of the SLHMC simulations with a larger system size (216 atoms) and compare those with experimental data. 
Radial distribution functions $g(r)$, the angle distribution $P(\theta)$, and structural properties of liquid silica obtained by SLHMC are summarized in Fig.\ref{Fig3}, \ref{Fig4} and Table \ref{tab1}. 
The calculated density of liquid silica was close to the experimental value 0.062 \cite{liquid_density1,liquid_density2} at 2373 K, 
and the density change via temperature was small.
Although the peak heights of $g(r)$ decreased as the temperature increase, 
the first peak positions of $g(r)$ were almost unchanged. 
The change of the first peak structures can be found as a tail toward a lower distance of Si-Si radial distribution at high temperature (see the inset of Fig.\ref{Fig3}). 
In the same manner, although the O-Si-O angle distribution shape became broad as the temperature increase, the average O-Si-O bond angles were close to 109${}^{\circ}$. 
This result means that SiO${}_{4}$ tetrahedral units were well maintained even at high temperatures.
The main structural difference via temperature change can be found in the Si-O-Si angle distribution.   
The average Si-O-Si angle became lower at high temperature and 
the Si-O-Si angle distribution spread to incorporate lower angles as shown in Fig.\ref{Fig4}(a).
The broadening of the Si-O-Si angle distribution is consistent with the previous results from shorter DFT-MD runs \cite{Sarnthein1995_prl,Sarnthein1995_prb,PhysRevB.86.054104}.
The Si-O-Si angle distribution around 90${}^{\circ}$ is due to a formation of edge-sharing SiO${}_{4}$ tetrahedrons (see the inset of Fig.\ref{Fig4}(b)). 
The tails of the first peak of Si-Si radial distribution are due to the contribution of the edge-sharing tetrahedrons, 
which causes the attraction of Si atoms. 
The edge-sharing tetrahedrons can be regarded as the defect of SiO${}_{4}$ tetrahedral units 
as a consequence of Si-O bond recombination at high temperatures. 
Although the broadening of the Si-O-Si angle distribution toward lower angles can be confirmed by MD with empirical  force field \cite{PhysRevLett.64.1955}, 
the formation of the edge-sharing tetrahedrons was not detected (see Appendix B). 
These results suggest that the DFT and empirical force field calculation result in different
defect structure of SiO${}_{4}$ tetrahedral units in liquid silica.

\begin{figure}[h]
\begin{center}
\includegraphics[width=0.8\linewidth]{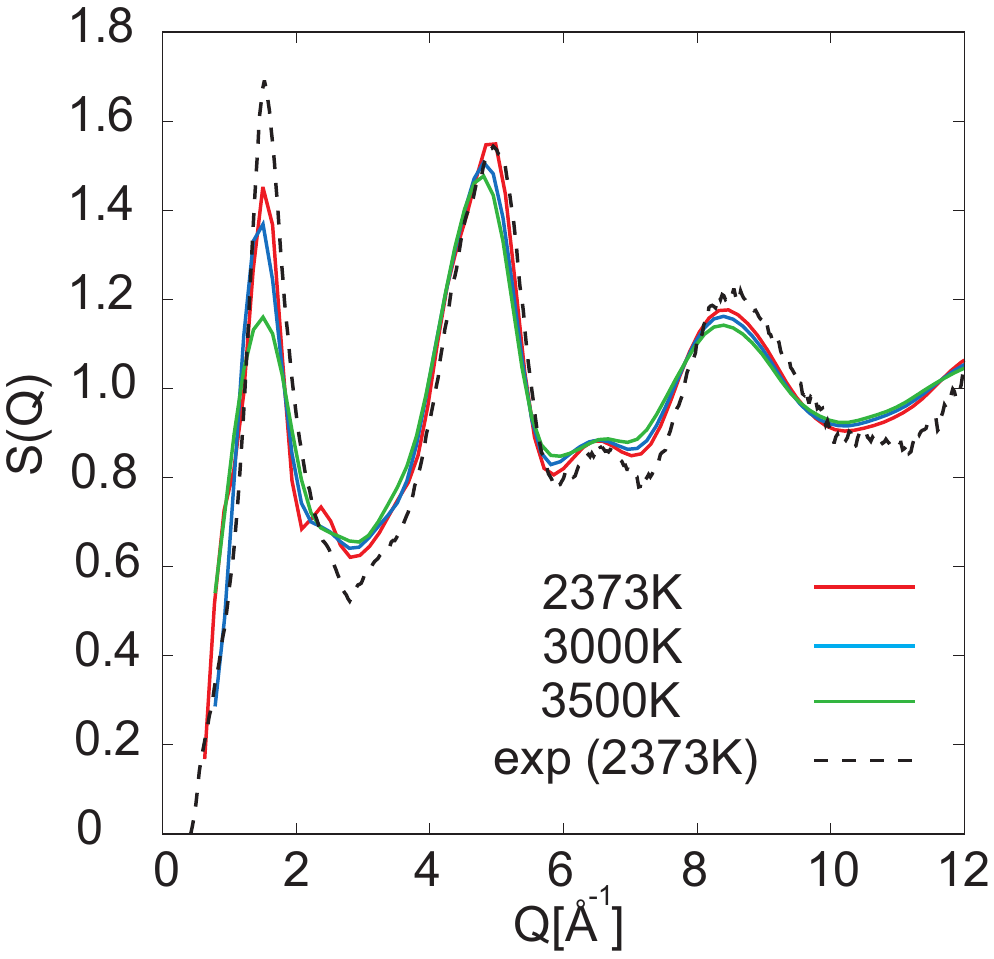}
\end{center}
\caption{
Total structure factor $S(Q)$ obtained by SLHMC and by the high-energy X-ray experiment \cite{Mei2007}.
}
\label{Fig5}
\end{figure}

We also calculated the total structure factor in X-ray diffraction. 
Using the Faber-Ziman partial structure factor \cite{Faber1965}
\be
S_{\alpha\beta}(Q) = 1+4\pi\rho_{0}\int drr^{2}\frac{\sin(Qr)}{Qr}(g_{\alpha\beta}(r)-1)\,,
\ee
total structure factor in X-ray diffraction can be calculated as 
\be
&&S_{\rm X}(Q)=\sum_{\alpha,\beta}\frac{c_{\alpha}c_{\beta}f_{\alpha}(Q)f_{\beta}(Q)}{
\langle f(Q)\rangle^{2}}S_{\alpha\beta}(Q)\,, \\
&&\langle f(Q)\rangle = \sum_{\alpha}c_{\alpha}f_{\alpha}(Q)\,,
\ee
where $\rho_{0}$ is the denstiy of liquid silica, 
$c_{\alpha}$ is the concentration of each species ($\alpha$=Si,O), and
$f_{\alpha}(Q)$ is the X-ray scattering factor for free ion \cite{Waasmaier1995}.
Figure \ref{Fig5} shows the structure factor obtained by SLHMC and high-energy X-ray experiment \cite{Mei2007}. 
%
%
Our SLHMC simulation accurately reproduces the experimental peak positions of $S(Q)$ except from the height of the first diffraction peak which is considered to be the finite size effect and is beyond the scope of this study. In fact, artifacts in the first peak was seen more clearly in the results of smaller systems with 72 atoms, as shown in Appendix C.

\section{CONCLUSION}
%
In this paper, we have developed the SLHMC method for isothermal-isobaric ensembles. 
This allows for an acceleration of first-principles DFT simulations of a soft material and liquids involving volume fluctuation. 
%
%
As a demonstration, we applied the isothermal-isobaric SLHMC method to the simulation of liquid silica at the temperature close to the experimental melting point.
We have shown that the SLHMC enables us to conduct very efficient sampling on the DFT-PES of liquid silica.
The MSDs obtained SLHMC developed much faster than the DFT-MD results with a small number of DFT calculations.
%
%
It is theoretically guaranteed that SLHMC reproduces all thermodynamic properties available from DFT-MD simulations, even though the approximate MLP is used.

The detailed structural properties of liquid silica were studied for the system with 216 atoms.
The obtained bond distances and O-Si-O bond angle were almost unchanged as the temperature increase. 
The main structural difference via temperature change was the broadening of Si-O-Si angle distributions toward lower angle, which was consistent with short DFT-MD runs previously reported \cite{Sarnthein1995_prl,Sarnthein1995_prb,PhysRevB.86.054104}.
The defect structure of SiO${}_{4}$ tetrahedral units at high temperatures was discussed.
We also calculated the structure factors and compared the results with the high-energy X-ray experimental data \cite{Mei2007}.
To the best of the authors’ knowledge, this is the first report of the fully first-principles calculation for the structure factor of liquid silica at the temperature close to the melting point.
%
The static structure factor obtained by SLHMC was in quite good agreement with the experimental data.

So far the SLHMC method has been developed based on the statistics of thermodynamic equilibria. 
To deal with rare event processes, the SLHMC might be suited to free energy calculations using biased sampling approaches. 
Combining the SLHMC with Wang-Landau \cite{PhysRevLett.86.2050}  and metadynamics methods \cite{Laio12562} would be interesting for a future work.

\section{SUPPLEMENTARY MATERIAL}
See supplementary material for the following: 
(1) accuracy of MLP in SLHMC, 
(2) computational efficiency of SLHMC.

\begin{acknowledgments}
M.S. thanks financial support from JSPS KAKENHI (18H05519, 18K05208, 21H01603) and MEXT Program for Promoting Researches on the Supercomputer Fugaku (Fugaku Battery \& Fuel Cell Project).
Y.N. thanks financial support from JSPS KAKENHI (20H05278).
The calculations were performed on the supercomputing system HPE SGI8600 at the Japan Atomic Energy Agency. 
The crystal structures were drawn with VESTA\,\cite{VESTA}.
\end{acknowledgments}

\section*{DATA AVAILABILITY}
The data that support the findings of this study are available from the corresponding author, K.K., upon reasonable request.

\appendix

\section{SLHMC and MLMD results at 4000 K}

\begin{figure}[h]
\centering
\includegraphics[width=1.0\linewidth]{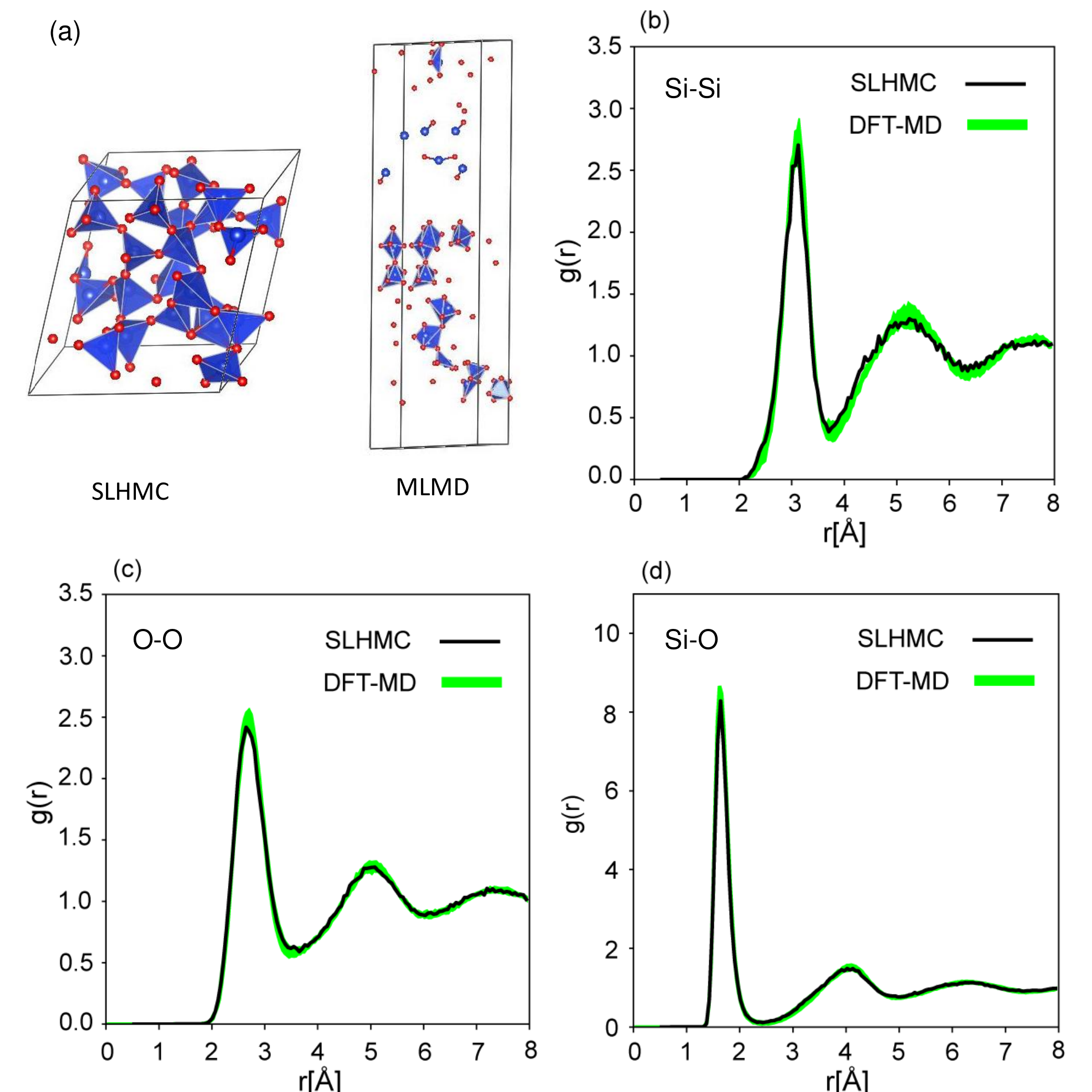}
\caption{(a):Final configulations of SLHMC (left) and MLMD (right) at 4000 K. (b), (c), and (d): Radial distribution functions for Si-Si, O-O, and Si-O pair obtained by SLHMC and DFT-MD at 4000 K}
\label{app1}
\end{figure}
In this appendix, we show the results obtained by SLHMC and machine learning molecular dynamics (MLMD) for liquid silica with 72 atoms at 4000 K. 
The training data set and the ANN architecture are the same in the main text. 
In SLHMC simulation, we conducted 10000 MCMC steps, and 
the same calculation conditions were used as in the main text. 
The resulting acceptance ratio and mean time interval of SLHMC were 19\% and 149 fs, respectively. 
In MLMD simulation, we adopted the No\`{s}e–Hoover thermostat and the Parrinello-Rahman barostat to generate NPT ensemble. 
The total simulation time of MLMD was 50 ps with 1 fs step size. 
Figure \ref{app1}(a) show the final configurations obtained by SLHMC and MLMD. 
The MLMD simulation shows structural collapse with a long simulation period. 
This result means that the MLP trained by 8600 training data and with the present ANN architecture is unstable 
and do not have accuracy to describe the diffusion dynamics at high temperature. 
On the other hand, the structural collapse was not confirmed in SLHMC simulation, 
since the SLHMC method rejects the MLMD trial move outside the DFT ensemble. 
The radial distribution functions calculated by SLHMC agree well with the results obtained by 80 ps long DFT-MD simulation as shown in Fig.\ref{app1}(b), (c), and (d). 
Thus, even if the MLP employed in SLHMC did not perfectly imitate the DFT-PES,
the SLHMC successfully generated the statistical ensemble on the DFT-PES. 

\section{Angle distributions obtained by MD with BKS potential}

\begin{figure}[h]
\centering
\includegraphics[width=1.0\linewidth]{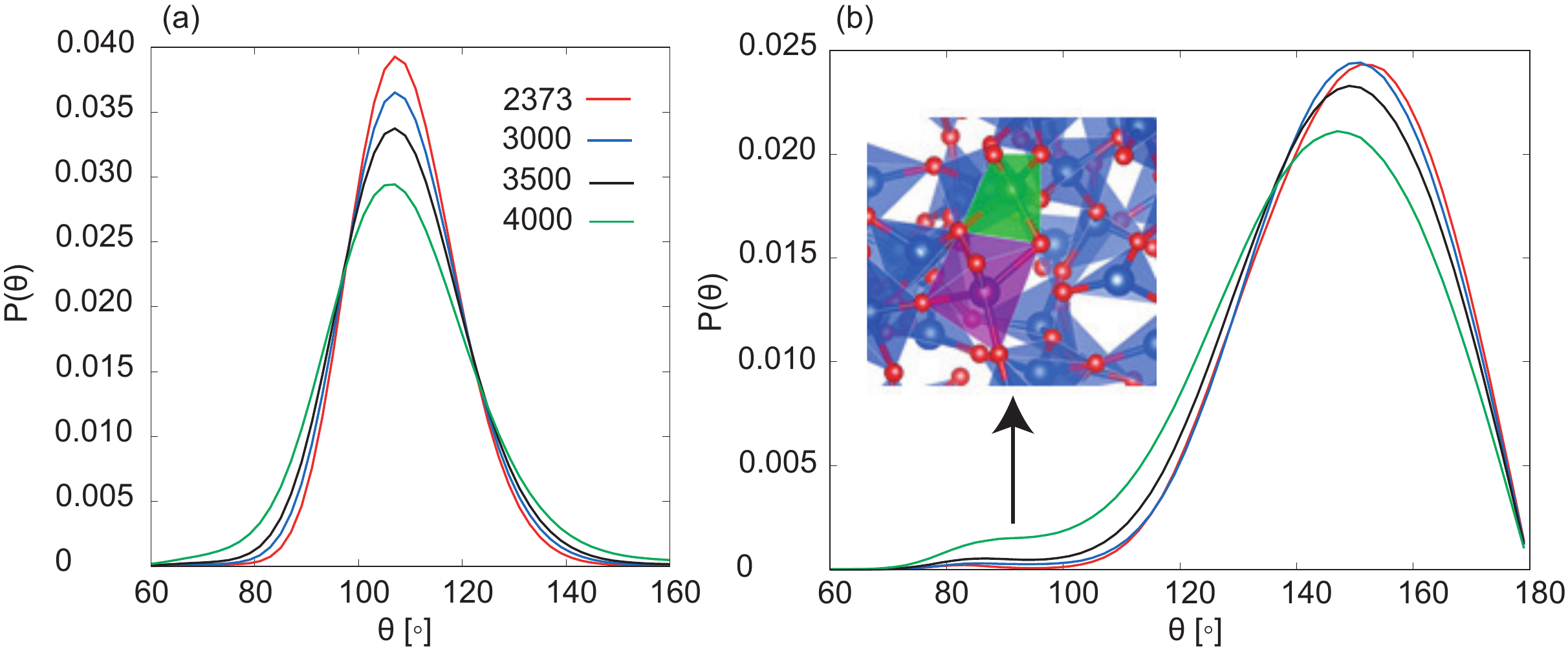}
\caption{
(a) and (b): Bond angle distribution $P(\theta)$ obtained by MD with BKS potential for the O-Si-O and Si-O-Si angles. 
The inset of Fig.(b) shows the edge-sharing tetrahedron and pentahedron 
(green and purple objects)
that contribute to the Si-O-Si angle distribution around 90${}^{\circ}$.
}
\label{app2}
\end{figure}
This appendix shows the bond angle distribution obtained by MD with 
the van Beest, Kramer, and van Santen (BKS) empirical potential \cite{PhysRevLett.64.1955} 
for liquid silica with 216 atoms.  
Non-Coulombic interatomic interactions of BKS were truncated at 5.5 {\AA} as done in reference \cite{Kob_1999}, and long-range Coulomb interactions were treated by a particle mesh Ewald method. 
The temperature and pressure were kept constant by using the Nosé-Hoover thermostat and 
the Parrinello-Rahman barostat, respectively. 
The total simulation time was 1 ns with 0.5 fs step size. 
Figure \ref{app2} shows the O-Si-O and Si-O-Si angle distribution $P(\theta)$ at 2373, 3000, 3500, and 4000 K.
The resulting average O-Si-O bond angles were close to 109${}^{\circ}$, indicating that 
SiO${}_{4}$ tetrahedral units were well maintained at high temperatures. 
The broadening of the Si-O-Si angle distribution toward lower angles was not observed until 
4000 K, while SLHMC results show the broadening of the angle distribution at 3000 K. 
The average Si-O-Si angles were reduced as the temperature increase as 
148.3, 147.7, 146.1, and 142.5${}^{\circ}$ at 2373, 3000, 3500, and 4000 K, respectively.
In the MD simulations with BKS potential, we could not confirm the edge-sharing SiO${}_{4}$ tetrahedrons in the trajectories.  
The Si-O-Si angle distribution around 90${}^{\circ}$ is due to a formation of the edge-sharing tetrahedron and pentahedron (see inset of Fig.\ref{app2}(b)).

\section{Structure factor $S(Q)$ with 72 atoms}

\begin{figure}[h]
\centering
\includegraphics[width=0.8\linewidth]{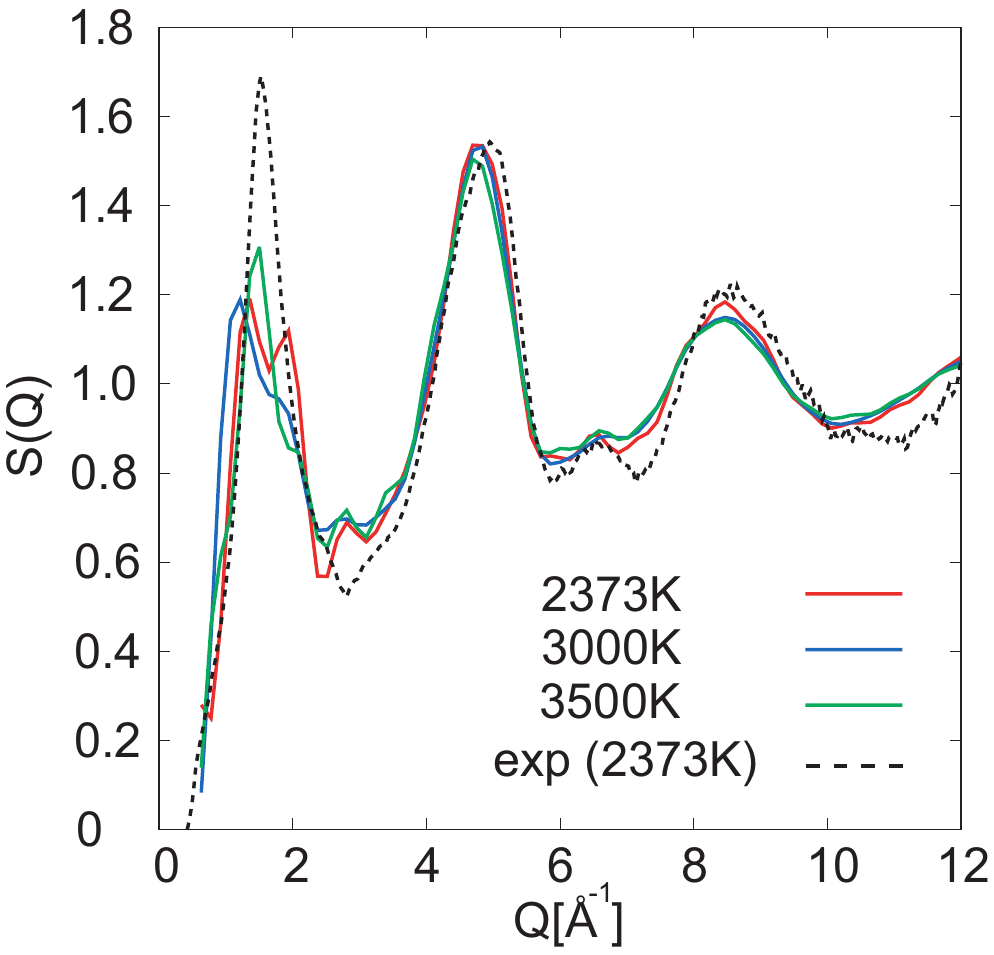}
\caption{Total structure factor $S(Q)$ obtained by SLHMC with 72 atoms and by the high-energy X-ray experiment \cite{Mei2007}.}
\label{app3}
\end{figure}
Figure \ref{app3} shows the structure factor obtained by SLHMC with 72 atoms and a high-energy X-ray experiment \cite{Mei2007}. 
Although the peak structures obtained by SLHMC above 4 \AA${}^{-1}$ agree well with the experimental data, the shape of $S(Q)$ with lower wave vector $(4 <Q)$ show large discrepancy between the SLHMC and experimental results due to the finite size effect.
The discrepancy of $S(Q)$ at lower wave vector were improved in SLHMC simulation for liquid silica with 216 atoms as shown in the main text. 

\nocite{*}

\bibliography{ref}

\end{document}